\documentclass[prl,aps,twocolumn,showpacs]{revtex4}
\usepackage{graphicx}
\usepackage{wasysym}
\usepackage{bm}
\usepackage{pstricks}

\newcommand{\beq}{\begin{equation}}
\newcommand{\eeq}{\end{equation}}

\newcommand{\bea}{\begin{eqnarray}}
\newcommand{\eea}{\end{eqnarray}}
\newcommand{\etal}{{\em et al.}}

\newcommand{\bq}{{\bf q}}
\newcommand{\bs}{{\bf S}}
\def\suhe#1{{\noindent\bf#1:}}
\def\pat#1{{\cal D}#1}

\def\tit#1#2#3#4#5{{#1}{\bf #2}, #3 (#4)}

\def\prep{Phys.\ Rep.\ }
\def\prl{Phys.\ Rev.\ Lett.\ }
\def\pr{Phys.\ Rev.\ }
\def\prb{Phys.\ Rev.\ B\ }

\def\jpco{J.\ Phys.\ Cond.\ Mat.\ }

\def\zpb{Z.\ Phys.\ B\ }

\def\natu{Nature\ }

\def\cjp{Can.\ J. Phys.\ }

\begin{document}

\title{Dipolar spin correlations in classical pyrochlore magnets}

\author{S. V. Isakov,$^1$ K. Gregor$^2$, R. Moessner$^3$,  and S. L. Sondhi$^2$}

\affiliation{$^1$Department of Physics, AlbaNova, Stockholm University,
SE-10691 Stockholm, Sweden} 
\affiliation{$^2$Department of Physics, Princeton University,
Princeton, NJ, USA}
\affiliation{$^3$Laboratoire de Physique
Th\'eorique de l'Ecole Normale Sup\'erieure, CNRS-UMR8549, Paris,
France}

\begin{abstract}
We study spin correlations for the highly frustrated
classical pyrochlore lattice antiferromagnets with $O(N)$ symmetry
in the limit $T \rightarrow 0$.
We conjecture that a local constraint obeyed by the extensively
degenerate ground states
dictates a dipolar form for the asymptotic spin correlations, at all
$N \ne 2$ for which the system is paramagnetic down to $T=0$. We verify
this conjecture in the cases $N=1$ and $N=3$ by simulations and to
all orders in the $1/N$ expansion about the solvable $N=\infty$ limit.
Remarkably, the 
$N=\infty$ formulae are an excellent fit, at all distances, to the 
correlators at $N=3$ and even at $N=1$.  Thus we obtain a simple 
analytical expression also for the correlations of the equivalent 
models of spin ice and cubic water ice, $I_h$. 
\end{abstract}

\maketitle

\suhe{Introduction}
In frustrated magnets, magnetic order is strongly suppressed 
compared to expectations 
from simple mean-field theory. This gives rise to a
cooperative paramagnetic, or spin liquid, regime, where the energy
scale of interactions exceeds that set by the temperature, and
nonetheless no long-range order ensues; if frustration is
particularly severe, magnetic ordering can be absent altogether. A
particular case in point is the pyrochlore lattice, where the absence
of ordering for the Ising antiferromagnet was already argued in
1956\cite{pyrowa}.
The Ising model is of particular interest on account of its $T=0$ entropy
and its equivalence to cubic ice $I_h$, and it has 
also been approximately realised experimentally in the
titanate spin ice compounds\cite{bramgingrev}.
There is strong evidence that the Heisenberg model on the 
pyrochlore lattice does not order down to 
$T=0$ \cite{villain,reimersmc,moecha} and a general theory based
on constraint counting indicates that this is so for all $O(N)$
magnets with $N>3$ although not for $N=2$, where thermal fluctuations
lead to collinear ordering (``order by disorder'')\cite{fn-plane,moecha}.

\begin{figure}[ht]
{
\centerline{\includegraphics[angle=0, width=2.5in]{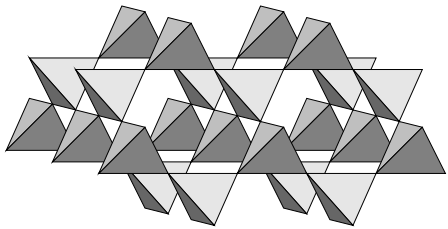}}
\caption{The pyrochlore lattice.}
\label{fig:pyrolat}}
\end{figure}

\begin{figure}[ht]
{
\centerline{\includegraphics[angle=0, height=2in,width=2.82in]{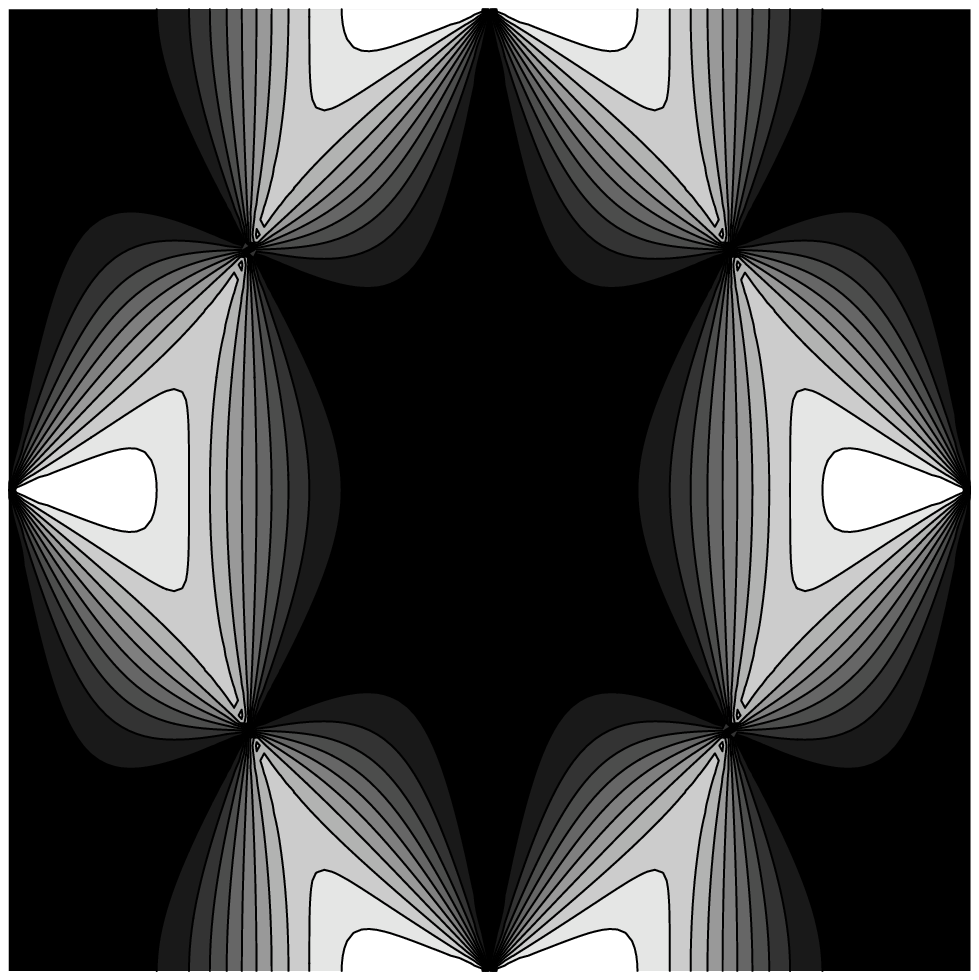}}
\centerline{\includegraphics[angle=0, width=2.8in]{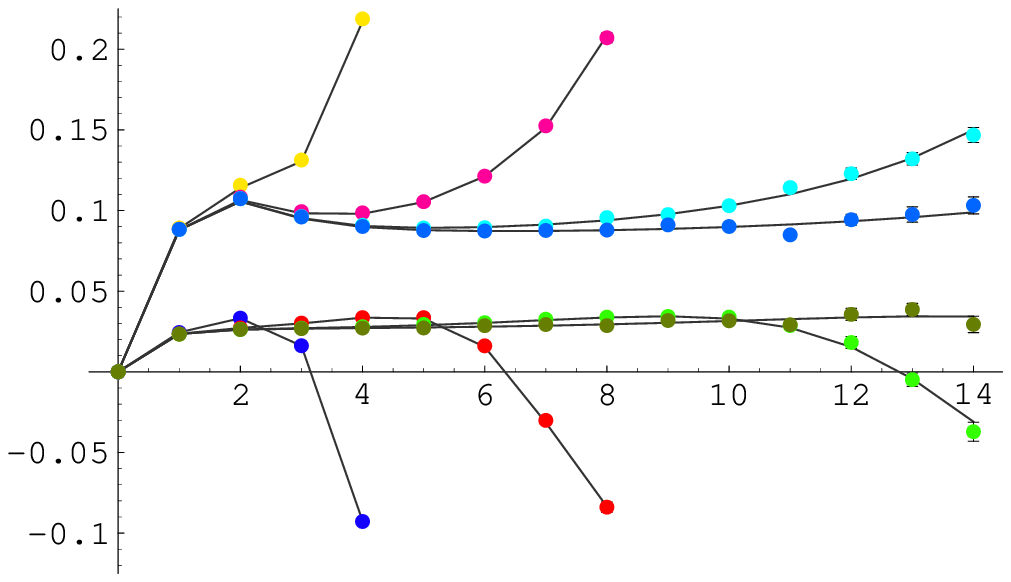}}
\centerline{\includegraphics[angle=0, width=2.8in]{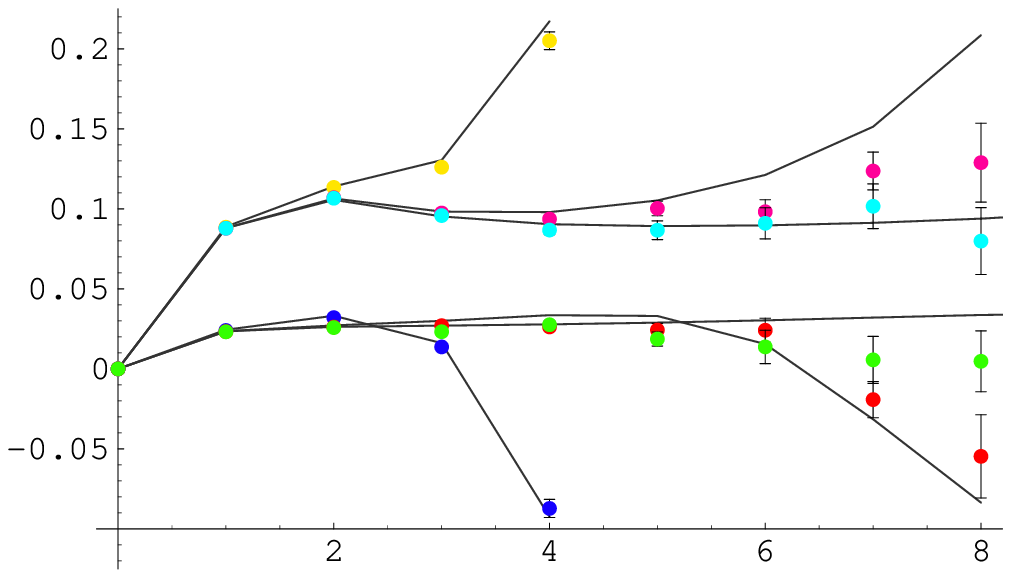}}
\caption{Top: The $[hhk]$
structure factor for $N=\infty$ [dark=low intensity; the x (y) axis corresponds to h (k)]. 
Middle: Ising correlations at $T=0$ 
(bottom: Heisenberg, at $T/J=0.005$) 
from Monte Carlo simulations, compared to $T=0$
large-$N$ correlations: $\langle
S_{1}(x) S_{1}(0)\rangle$, for two inequivalent directions,
$[101]$ and $[211]$,
multiplied by
the cube of the distance, $x^3$. The error bars are around
$5\times 10^{-6} x^3$ ($5 \times 10^{-5} x^3$). Different curves
are for system sizes $L=8, 16, 32, 48$ ($L=8, 16, 32$).
There is no fitting parameter.}
\label{fig:pyrocorr}}
\end{figure}

The absence of long range order does not mean the physics is
trivial at low temperatures as the accessible states have
non-trivial local constraints on them---hence {\it cooperative}
paramagnet.
For the thermodynamics it is possible to make do with small
clusters, e.g. Pauling's estimate of the entropy of ice
works rather well and
the thermodynamics of the Heisenberg pyrochlore magnet is
well-described even by an approximation based on an {\em isolated}
tetrahedron\cite{moeberl}.
For the spin correlations however, the situation is presumably
different. Indeed, Monte Carlo simulations of the Heisenberg magnet
by Zinkin \etal\cite{zinkinmc} 
have demonstrated the presence of sharp
features in the structure factor in certain high symmetry directions
(the ``bow-ties'' in Fig.~\ref{fig:pyrocorr}), indicative of extended
spin correlations. It is understood, qualitatively, that
these are a direct consequence of the ground state constraint that each
tetrahedron separately has vanishing total spin\cite{moecha}.

In this paper we provide a theory of the $T \rightarrow 0$
spin correlations of
nearest neighbor $O(N)$ magnets on the pyrochlore lattice when they are
paramagnetic down to $T=0$. This restriction excludes only the case
$N = 2$, but even there our results are probably applicable as we discuss
below.  Our work builds on two previous advances.  The first is the
recognition\cite{youngaxe3d,clhu,HKMS3ddimer,hermele03}, generalizing work on
two-dimensional ice\cite{youngaxe}, that $d=3$ models with binary degrees of
freedom and a local ``Gauss's law'' constraint should generally exhibit
asymptotic dipolar
correlations governed by a pure Maxwell action. This applies directly to our
$N=1$ (Ising) case, while the underlying argument has been conjectured to
apply also to Heisenberg spins by Henley\cite{clhu}.  The second is the
solution by Garanin and Canals of the $N=\infty$ limit \cite{canalsgaranin}, 
in particular their numerical determination of
the structure factor in the $N=\infty$ limit, which exhibits the same bow-ties
observed by Zinkin \etal

In the following we (a) motivate how dipolar correlations arise in 
pyrochlore magnets from local constraints,
(b) demonstrate their existence analytically in the $N=\infty$ limit
and their persistence to all orders in the $1/N$ expansion, and
(c) verify by simulation that the correlations at $N=3$ and $N=1$
are dipolar.
We find that the correlations at $N=3$ are very well fitted by the 
$N=\infty$ formulae even without modifications at short distances. 
Remarkably we find that the $N=\infty$ formulae are probably even
more accurate for $N=1$ where the relative and absolute error never
exceed $2\%$ and $4.5\times10^{-4}$, respectively even though this is not
a conclusion that one would guess on the basis of the $1/N$ expansion!
We close by noting the applicability of the ideas in the paper to
related systems, which will be detailed in  a separate
publication\cite{imswip}, as well as their known limitations.

\suhe{Local constraints and dipolar correlations} We consider the
classical nearest neighbor antiferromagnet on the pyrochlore lattice,
$H=J\sum_{\langle i,j\rangle} \bs_i\bs_j$, where the $ \bs_i$ are 
$N$-component spins of fixed length $N$. We set $J=1$ in the following.
The pyrochlore lattice consists
of corner sharing tetrahedra, and the Hamiltonian can be rewritten, up
to a constant, as $H=(1/2)\sum_{\XBox}(\sum_{i \epsilon \XBox}\bs_i)^2$,
where the sum in parentheses runs over all four spins at the corners
of a given tetrahedron, $\XBox$, and the outer sum is over all tetrahedra. 
Hence, the ground states (minimum energy
configurations) are such that for each tetrahedron
and each spin component $\alpha$, 
\bea 
\sum_{i \in \XBox}
S_i^\alpha = 0.
\label{eq:gsconstraint}
\eea

This can be turned into a manifest conservation law on the dual --
{\em bipartite} diamond -- lattice, the sites of which are given by
the centres of the tetrahedra while the spins sit at the midpoints of
its bonds. 

\begin{figure}[ht]
{
\centerline{\includegraphics[angle=0, width=2.0in]{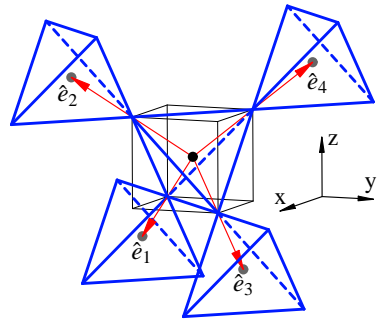}}
\caption{The centres of the tetrahedra define (the two sublattices of)
the diamond lattice, denoted by black and grey dots. The orientation
of the diamond bonds, and the four sublattices of the pyrochlore
lattice, are defined by the vectors $\hat{e}_\kappa$}.
\label{fig:pyroorient}}
\end{figure}

First, we orient each bond, $\kappa$, by defining a unit vector
$\hat{e}_\kappa$, which points along the bond from one sublattice to
the other, see Fig.~\ref{fig:pyroorient}. Next we define $N$ vector fields
on each bond, ${\bf B}^\alpha_\kappa = S^\alpha_\kappa \hat e_\kappa$, where
$S^\alpha_\kappa$ denotes the spin on bond $\kappa$.  The ground state
constraint (\ref{eq:gsconstraint}) implies that each ${\bf B}^\alpha$
separately forms a set of solenoidal fields at $T=0$, $\nabla \cdot
{\bf B}^\alpha =0$. 

For $N=1$, spin flips connecting two ground states correspond to
reversing the direction of a closed loop of ``magnetic flux'', ${\bf
  B}$; evidently, ${\bf B}$ averages to zero over such a flippable
cluster of spins. Upon coarse-graining, a high density of flippable
clusters (and therefore a large number of ground states) translates
into a small (well-averaged) coarse-grained ${\bf \tilde{B}}$. We now
posit that this feature carries through to $N>1$, so that states with
small values of ${\bf \tilde{B}^\alpha}$ will in general be
(entropically) favoured. This is captured most simply by introducing
a weight functional, $\rho$:  
\bea 
\rho[\{{\bf \tilde{B}}^\alpha ({\bf
  x})\}] \propto \exp \left[-\frac{K}{2}\int d^3 x \sum_\alpha 
({\bf \tilde{B}}^\alpha)^2
\right]
\label{eq:gaugeansatz}
\eea 
provided the solenoidal constraint is implemented, where $K$ is the
stiffness constant.
(If we solve the constaint by introducing a vector
potential for each component, we are led to the Maxwell action.) 

From this we can deduce the long distance correlators, 
\beq  \label{spacedipole}
\langle
\tilde{B}^\alpha_i({\bf x}) 
\tilde{B}^\beta_j(0) \rangle \propto \delta_{\alpha\beta}
{3 x_i x_j - r^2 \delta_{ij}\over r^5} 
\eeq 
which are dipolar as
advertised. The appropriate long-wavelenth formulae for the case $N=1$
have already been given in Ref.~\onlinecite{hermele03}.

This argument does not take into account thermal fluctuations {\em out
  of} the ground state manifold. These are gapped, and thus
unimportant for $T\rightarrow 0$, for $N=1$. However, for continuous
spins they endow each microscopic ground state with a non-trivial entropic
weight, and are essential for even defining what is meant by a
measure on the set of ground states. An analysis based on Maxwellian
constraint counting implies that this weighting is sufficiently uniform to be
innocuous for $N\geq 3$ so that the dipolar forms should hold. 
For $N=2$ this is not the case and the weighting
leads to an order-by-disorder phase transition \cite{moecha,fn-plane}.  
However at this transition, the spins order collinearly but do not
pick fixed orientations along their common axis.
Hence, for the resulting collinear ensemble, the $N=1$ results can be
applied, with the caveat that it has not been established whether a
further phase transition will select a subset of collinear states at
even lower temperatures.

While the arguments given above are based on the plausible ansatz
(\ref{eq:gaugeansatz}), we now turn to establishing their actual
correctness. The continuity of the physics for $N > 3$ will allow
us to tackle this in the $1/N$ expansion and we will supplement this
with explicit simulation at small $N$.

\suhe{$\bm N=\infty$} We start with the simple classical $O(N)$
model; its $N=\infty$ saddle point\cite{moshemoshe} 
is, of course, the one discussed in
Refs.~\onlinecite{canalsgaranin}.  
The partition function is given by $Z=\int
\pat{\phi}\pat{\lambda} \exp[-{\cal  S(\phi,\lambda)}]$, 
with the action defined as
\bea
{\cal S(\phi,\lambda)}&=&\sum_{i,j}
\sum_{\alpha=1}^N
%\left\{ 
\frac{1}{2}
\phi_i^\alpha A_{ij} \phi_j^\alpha+
i\frac{\lambda_i}{2}\delta_{i,j}(\phi_i^\alpha \phi_i^\alpha-N)\ ,
%\right\}
%\nonumber\\
%&\equiv& 
%\sum_{i,j}
%\sum_{\alpha=1}^N
%\left\{ \frac{1}{2}
%\phi_i^\alpha M_{ij} \phi_j^\alpha-i N
%\frac{\lambda_i}{2}\delta_{i,j}
%\right\}
%\ ,
%\label{eq:saddact}
\nonumber
\eea
where $A_{ij}$ is the interaction matrix divided by $T$.

The saddle point evaluation for $N=\infty$ is analogous to that of
Refs.~\onlinecite{canalsgaranin}; 
as we are not aware of an analytical
treatment of the full pyrochlore correlations in the literature, we provide
it here.

The Fourier transform of the interaction matrix is 
\bea
  2 \left( \matrix{ 1 & c_{xz} & c_{xy} & c_{yz} \cr 
   c_{xz} & 1 & c_{\overline{yz}} & c_{\overline{xy}} \cr 
   c_{xy} & c_{\overline{yz}} & 1 & c_{\overline{xz}} \cr
   c_{yz} & c_{\overline{xy}} & c_{\overline{xz}} & 1 } \right),
\label{eq:adj}
\eea
where $c_{ab}=\cos(\frac{{q_a}+{q_b}}{4})$ and
$c_{\overline{ab}}=\cos(\frac{{q_a}-{q_b}}{4})$.
%$c_{yz}=\cos(\frac{{q_y}+{q_z}}{4})$,
%$c_{xz}=\cos(\frac{{q_x}+{q_z}}{4})$, $c_{xy}=\cos(\frac{{q_x}+{q_y}}{4})$,
%$c_{\overline{xy}}=\cos(\frac{{q_x}-{q_y}}{4})$,
%$c_{\overline{xz}}=\cos(\frac{{q_x}-{q_z}}{4})$,
%$c_{\overline{yz}}=\cos(\frac{{q_y}-{q_z}}{4})$.
The eigenvalues of the interaction matrix are \cite{reimersberlchi}
\bea
\nu_{1,2}=0, \
\nu_{3,4}=2\pm\sqrt{1+Q} \nonumber,
\eea
where $Q=c_{xy}^2+c_{\overline{xy}}^2+c_{xz}^2+c_{\overline{xz}}^2
 +c_{yz}^2+c_{\overline{yz}}^2-3$.

These can be combined to give the spin correlators\cite{canalsgaranin}. 
For the zero
temperature case discussed here, only the modes with $\nu=0$
contribute. There are two independent
correlators, namely those between spins on the same and between 
spins on different
sublattices. We thus find the following spin correlators
(we label four sublattices as shown in Fig.~\ref{fig:pyroorient})
\bea
\langle S_1^\alpha (\bq)S_1^\beta (-\bq)\rangle
&=&\delta_{\alpha \beta}
\frac{6-
   2c_{\overline{xy}}^2-2c_{\overline{xz}}^2-2c_{\overline{yz}}^2}{3-Q},
\nonumber \\
\langle S_1^\alpha (\bq)S_2^\beta (-\bq)\rangle
&=& \delta_{\alpha \beta}
 \frac{2\left( \cos(\frac{{q_y}}{2})c_{\overline{xz}}
    - c_{xz}\right)}{3-Q} \ .
\label{eq:s11}
\eea
From these one can calculate the correlators of ${\bf B}^\alpha = 
\sum_\kappa S^\alpha_\kappa \hat e_\kappa$ and verify that they have
the small ${\bf q}$ forms,
\beq  \label{qdipole}
\langle
\tilde{B}^\alpha_i({\bf q})
\tilde{B}^\beta_j({\bf q}) \rangle \propto 
\delta_{\alpha\beta}
( \delta_{ij} - \frac{q_i q_j}{q^2} )
\eeq
equivalent to the real space forms (\ref{spacedipole}) noted earlier.

The structure factor in this limit is
\bea
&&S(\bq) = 4-\frac{8}{3-Q} \biggl[
      \left(c_{yz}+c_{\overline{yz}}\right){\sin^2\left(\frac{{q_x}}{4}\right)}
           \nonumber \\ 
  &+& \left(c_{xz}+c_{\overline{xz}}\right){\sin^2\left(\frac{{q_y}}{4}\right)}
   +  \left(c_{xy}+c_{\overline{xy}}\right){\sin^2\left(\frac{{q_z}}{4}\right)}
                           \biggr].
\label{eq:strfct}
\eea
It has been customary to consider the structure factor of the
pyrochlore antiferromagnets in the $[hhk]$ plane in reciprocal space,
as it contains the high-symmetry directions 
$[100]$, $[110]$ and $[111]$.
Here, the wavevector ${\bf q}=(q_x,q_x,q_z)$, and
Eq.~\ref{eq:strfct} simplifies to
\bea
S_{[hhk]}&=& \frac{32 {\left( \cos (\frac{{q_x}}{4}) - 
    \cos (\frac{{q_z}}{4}) \right) }^2 {\sin (\frac{{q_x}}{4})}^2}
    {5 - \cos ({q_x}) - 4 \cos (\frac{{q_x}}{2}) \cos (\frac{{q_z}}{2})}.
\eea
This is plotted in Fig.~\ref{fig:pyrocorr}; the same plot has
previously appeared in Ref.~\onlinecite{canalsgaranin}.  We emphasize
that it is the transverse, dipolar form of the correlators which gives
rise to the bow tie structures. 

\suhe{$1/N$ expansion} To set up the $1/N$ expansion \cite{moshemoshe}, 
we expand away from
the saddle point by allowing the Lagrange multiplier field $\lambda$
to vary from its uniform saddle point value, $\tilde\lambda$: 
$\lambda_i=\tilde\lambda+\mu_i$. Integrating out the $\phi$'s removes the term linear 
in $\mu$; the quadratic term yields the propagator and higher order terms generate vertices.
The quadratic term, $\sum_{ij}\mu_i \langle \phi_i
\phi_j\rangle^2\mu_j\equiv\sum_{ij}\mu_i {\cal M}^{-1}_{ij}\mu_j$,
implies a propagator $\langle \mu_i \mu_j \rangle = {\cal M}_{ij}$.
One can argue, and we have checked by explicit numerical computation,
that $\langle \mu_i \mu_j \rangle$ decays inversely as the sixth
power of the separation between $i$ and $j$. Hence its Fourier
transforms, ${\cal M}_{\kappa\kappa'}({\bf q})$, are continuous 
functions over the Brillouin zone. We note
that the $N=\infty$ correlators of the spins $G^0_{\kappa\kappa'}
({\bf q})$ are discontinuous at ${\bf q} =0$ on account of their
dipolar character.

We now briefly sketch the proof that the dipolar form on the spin
correlations survives to all orders in the $1/N$ expansion; details
will be presented in \cite{imswip}. Consider the Dyson series for
the spin correlator, $G$, at finite N:
\bea \label{dyson}
& & G_{\kappa \kappa'}({\bf q}) = G^0_{\kappa \kappa'}({\bf q})
+ G^0_{\kappa \kappa_1}({\bf q}) 
\Sigma_{\kappa_1 \kappa_2}({\bf q}) G^0_{\kappa_2 \kappa'}({\bf q}) 
+ \cdots\nonumber 
%\\
%& &
%G^0_{\kappa \kappa_1}({\bf q}) 
%\Sigma_{\kappa_1 \kappa_2}({\bf q}) G^0_{\kappa_2 \kappa_3}({\bf q})
 %\Sigma_{\kappa_3 \kappa_4}({\bf q}) 
 %G^0_{\kappa_4 \kappa'}({\bf q}) + \cdots
\eea
where the self energy $\Sigma$ is the sum of the relevant graphs at
all orders in $1/N$.
For each graph that enters this sum one can show (I) that 
$\Sigma_{kl}({\bf q})$ is a continuous function of $q$ in the Brillouin zone, 
(II) that $\lim_{\bf q \rightarrow 0} \Sigma_{\kappa \kappa'}({\bf q})
= \sigma^{diag}$ for $\kappa=\kappa^\prime$ and $\sigma^{od}$ otherwise.
(I) follows from the fact
that in all such diagrams one integrates over an integrand that
is bounded and 
at most discontinuous at isolated points, while (II)
uses the action of lattice symmetries on $G^0$
and ${\cal M}$.
Using these observations the leading small $q$ terms in the Dyson
series  lead to a simple geometric sum and one finds that
\beq
G_{\kappa \kappa'}({\bf q})\propto G^0_{\kappa
\kappa'} ({\bf q})
\eeq
near ${\bf q} =0$. Consequently,
the correlations are still dipolar at long distances. 
%Their coefficient is modified -- 
%but only weakly: explicit evaluation of the leading corrections yields $G\sim(1-0.03/N)G^0$.
%Nonetheless, even in
However, even  in 
the most optimistic case we would not expect the series  to converge 
beyond $N=3$, where long range order sets in due to order by
disorder. 

\suhe{Comparison to Monte Carlo simulations} To the naked eye, the
$N=\infty$ structure factor of Fig.~\ref{fig:pyrocorr} looks almost identical to
the one obtained by Zinkin for Heisenberg spins\cite{zinkinmc}
and by us for Ising spins\cite{imswip}.
For a  more sensitive test we have compared the real-space
expressions to Monte Carlo simulations for Ising and Heisenberg spins
for different system sizes, shown in the same figure.  We find that
the error for Ising spins is very small indeed -- the relative and absolute
systematic 
error never seems to exceed $2\%$ and $4.5\times10^{-4}$ respectively.
The agreement for
Heisenberg spins is somewhat worse. However, since the disagreement is
of the size of the finite-temperature error bars, we cannot
extrapolate down to $T \rightarrow 0$ reliably. In either case, one
sees that the $N=\infty$ formulae we have derived work remarkably well
quantitatively even without any further corrections---indeed, far better 
than one might have guessed at the outset---even at small distances, 
and they correctly capture finite-size effects.  

\suhe{Applicability to other models} 
As the Ising antiferromagnet on the pyrochlore lattice is isomorphic to
(nearest-neighbour)\cite{enjgin} 
spin ice, the results can straightforwardly
be transcribed, and further carried over to cubic (water) ice
$I_h$, which is again equivalent\cite{pyrowa}. In the latter context the
asymptotic form of the correlations is known \cite{youngaxe3d} and the
interest of our results is in the accuracy of our analytic forms
at all distances.
The analysis presented here
straightforwardly generalises to other three dimensional cooperative
paramagnets with conservation laws.

The present approach can also be applied to two-dimensional models
with a conservation law. However, it breaks down for models with a
discrete set of ground states. Technically, the effective height
(two-dimensional gauge) field acquires a non-zero compactification
radius in this case, which leads to the appearance of additional
operators under coarse-graining. These are absent for $N=\infty$, and
their appearance is non-perturbative in $N$. For the kagome magnet,
for example, they arise for $N=2,3$, which is why the
$\sqrt{3}\times\sqrt{3}$ correlations known to be present there were
not found in Ref.~\onlinecite{canalsgaranin}. However, information
gleaned from the present approach, such as a stiffness,  
can nonetheless usefully be fed into
these models. Finally, we have shown that the present approach can
also be applied to a class of paramagnetic models without a conservation 
law, for
which so far very little has been known in terms of correlation
functions. This, along with the more technical material, will be
discussed elsewhere\cite{imswip}.

\suhe{Acknowledgements}
We would like to thank Hans Hansson, Mike Hermele, Chris Henley,
Anders Karlhede, Oleg Tchernyshyov and Kay Wiese for useful
discussions. We are also grateful to David Huse and Werner Krauth for
collaboration on closely related work.  This work was in part
supported by the Minist\`ere de la Recherche et des Nouvelles
Technologies with an ACI grant, by the NSF with grants DMR-9978074 and
0213706, and by the David and Lucile Packard Foundation.

\end{document}